\documentclass[aps,prl,twocolumn, showpacs]{revtex4}

\usepackage{graphicx}
\usepackage{dcolumn}
\usepackage{amsmath}
\newcommand{\gamnas}{${\rm Ga}_{1-x}{\rm Mn}_{x}{\rm As}$ }
\newcommand{\be}{\begin{equation}}
\newcommand{\ee}{\end{equation}}
\newcommand{\comma}{,\;}

\begin{document}

\title{Scaling analysis of the magnetoresistance in \gamnas:
Evidence for strong fluctuation and interaction effects}
%\author{AUTHORS}
\author{C. P.  Moca$^{1,2}$, B. L. Sheu$^3$, N. Samarth$^3$, P. Schiffer$^3$, B. Janko$^{4,5}$, and G. Zarand$^{1}$}
\affiliation{
$^1$ Budapest University of Technology and Economics\comma H-1521 Budapest\comma Hungary\\
$^2$ Department of Physics\comma University of Oradea\comma 410087 Oradea\comma Romania \\
$^3$ Department of Physics and Materials Research Institute\comma Pennsylvania State University\comma University Park\comma PA 16802\comma USA \\
$^4$ Department of Physics\comma University of Notre Dame\comma
Notre Dame\comma Indiana\comma 46556\comma USA\\
$^5$ Materials Science Division\comma Argonne National
Laboratory\comma 9700 South Cass Avenue\comma Argonne\comma
Illinois\comma 60439\comma USA}

\date{\today}
\begin{abstract}
We compare  experimental resistivity data on \gamnas films with
theoretical calculations using a scaling theory for strongly
disordered ferromagnets. All  characteristic features of the
temperature dependence of the resistivity can be quantitatively
understood through this approach as originating from the close
vicinity of the metal-insulator transition. In particular, we find
that the magnetic field induced changes in resistance cannot be
explained within a mean-field treatment of the magnetic state, and
that accounting for thermal fluctuations is crucial for a
quantitative analysis. Similarly, while the non-interacting scaling
theory is in reasonable agreement with the data, we find clear
evidence in favor of interaction effects at low temperatures.
%localization effects in the presence of electron-electron interaction.
\end{abstract}
\pacs{72.20.Ði, 72.80.Ðr, 75.50.Pp, 72.20.My}
\maketitle

Over the past decade, \gamnas has been the most extensively studied
ferromagnetic semiconductor because it can be used in
proof-of-concept spintronic devices~\cite{AMacDonald}. In
particular, the carrier-mediated ferromagnetism in
this material makes it attractive for device integration with the technologically mature
III-V semiconductors such as GaAs, where bandgap engineering allows systematic
modulation of the carrier density in heterostructure devices. Furthermore, we
note that an all-electrical III-V semiconductor spintronic device has been
recently demonstrated~\cite{Lou}.  It is important within this general context
to develop a fundamental
understanding of the interplay between carrier transport and magnetism in
\gamnas.  One of the basic
but least understood properties of \gamnas~ is the temperature and
magnetic field dependence of its resistivity~\cite{matsukura_PRB_98}.

\begin{figure}[b]
\centering
\includegraphics[width=3.3in]{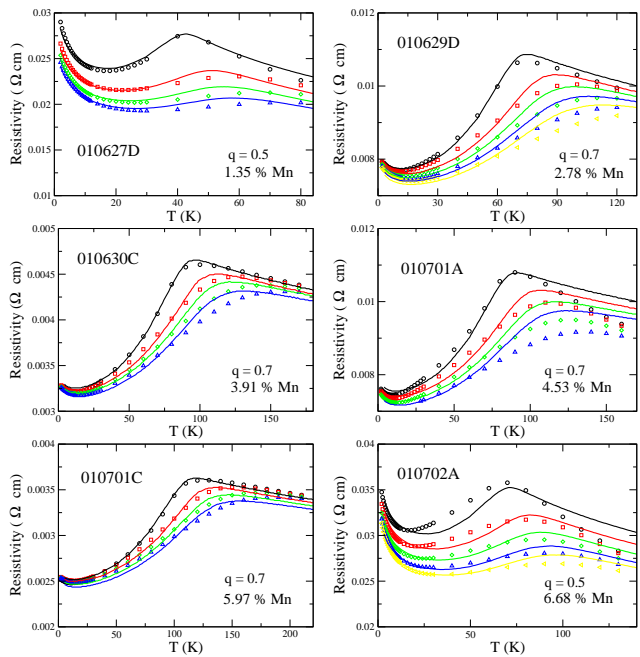}
\caption{Comparison between the experimental data and the
theoretical results at magnetic fields H =0, 3, 6 and 9 T. Dots
represent experimental data,  solid lines are theoretical
fits. Each figure indicates the value of $q$' that is used for
theoretical fitting.} \label{fig:resistivity}
\end{figure}

For typical \gamnas samples the resistivity increases with
decreasing temperature above the Curie temperature, $T_C$, but then
it suddenly drops below $T_C$ thereby resulting in a resistivity
peak at $T_C$, which then gradually broadens and gets shifted to
higher temperatures  upon application of a magnetic field (see
Fig.~\ref{fig:resistivity}).  In ``high quality samples'', the
aforementioned increase is less pronounced and one observes a broad
shoulder rather then a resistivity peak. The resistivity  also shows
another upturn at much lower temperatures. There have been a number
of attempts to explain the resistivity peak so
far~\cite{matsukura_PRB_98}, which invoked, e.g., scattering by
critical fluctuations~\cite{Fisher},  the formation of magnetic
polarons~\cite{Littlewood,Dietl}, 'dynamical' mean field
calculations~\cite{dasSarma}, or the interplay with universal
conductance fluctuations~\cite{Felix}. Although these theoretical
approaches have been quite successful in addressing a particular
range or qualitative aspect of the data, however, a theoretical
framework that could quantitatively explain all characteristic
features observed in \gamnas has not been available so far. Besides
the simplicity of our model and the clear set of reasonable
parameter choices we make, the attractive feature of the approach
presented in this paper stems from the fact that it is capable of
{\em quantitatively} account for all distinct features of the
experimental data: $(i)$ the gradually increasing resistance as the
temperature is lowered towards $T_C$; $(ii)$ the pronounced peak
precisely at $T_C$; $(iii)$ the upturn in resistance at low
temperatures, together with the finite resistance intercept for
metallic samples; $(iv)$ the precise amount with which an external
magnetic field depletes the resistance peak at $T_C$ and shifts it
towards higher temperatures; $(v)$ the remarkable "non-crossing"
constraint of the experimental data, a constraint that some other
theories fail: the resistance curves that belong to different values
of the external field do {\it not} cross.

In this Letter, we present a theoretical analysis of the temperature
dependent resistivity of \gamnas and show good agreement with a
range of measured samples despite of the simple assumptions we make.
First of all, we remark that the features mentioned above are
reminiscent of localization effects interplaying with magnetism.  In
fact, most $III-V$ ferromagnetic semiconductors are very bad
conductors because the charged dopants (Mn for \gamnas) introduce
very large disorder~\cite{Carsten} and can lead even to the
formation of an impurity band~\cite{impurity_band}. As we show
later,  even for annealed and relatively highly-doped samples, a
naive estimate gives $k_F l \sim  0.3$, with $k_F$ the Fermi
momentum and $l$ the mean free path. This value clearly hints
towards the importance of disorder, and suggests that it is
necessary to beyond the popular free electron picture frequently
used in the literature~\cite{Review}, . \cite{Basov}.
Strikingly similar resistivity anomalies have been also observed in
other types of magnetic  semiconductors
\cite{anomaly_semiconductors}, as well as some manganites
\cite{anomaly_manganites,Viret}, with various semi-phenomenological
frameworks available for explaining these phenomena in terms of
localization theory \cite{Viret,Kogan,Alascio}. However, these
approaches focus explicitly on the localized phase. Such a starting
point is not suitable for a fairly large class of experimentally
measured {\gamnas } samples that are not insulators, but very poor
metals, close to the localization transition.

In Ref.~[\onlinecite{Gergely}], we developed a scaling
theory of magneto-resistance, wherein all the
characteristic features of the resistivity anomalies appear
naturally. This previous work primarily focused on the localized
phase, where Mott's variable range hopping formula can be used to
compute the resistivity.
The purpose of the present paper is to extend  this theory
to  metallic samples, and show that it {\em
qualitatively} and {\em quantitatively} explains detailed aspects of
the magneto-transport properties. To understand the properties
of \gamnas we  must take into
account electron-electron interaction too. Nevertheless, as explained later,
 the concept of
universal one-parameter scaling carries over
to this material due to the large intrinsic spin-orbit coupling of \gamnas
and the presence of magnetic moments.

To test the theoretical approach, we have measured a series
of annealed and unannealed \gamnas samples grown by molecular beam epitaxy in which
the Mn concentration (x) was systematically varied between 0.0135 and
0.067. The temperature dependent resistivity data shown in
Fig.~\ref{fig:resistivity}
were measured in different magnetic fields inside a commercial
cryostat (Quantum Design PPMS), with the magnetic field  normal
to the sample plane. Details of the sample growth, materials
characterization and annealing protocol were reported
elsewhere~\cite{Potashnik}.

%
%The theory of Ref.~\onlinecite{Gergely} is a natural extension of
%the scaling theory of Abrahams {\it et al.} \cite{Abrahams},
%constructed to describe the disorder-induced metal-insulator transition.
In the present paper we shall use a slightly  modified version of the
scaling approach applied for disordered
conductors.\cite{Abrahams,Finkelstein,Castellani}
In Ref.~\cite{Gergely} we considered non-interacting electrons, where a
scaling theory can be constructed in terms of the dimensionless conductance, $g$
and a lengthscale at which electrons loose their coherence.
In the presence of interactions, one also needs to introduce
the dimensionless interaction parameters in the triplet and singlet channels,
$\gamma_t$ and $\gamma_s$, respectively.
 However, for  \gamnas an important simplification occurs:
\gamnas has a very large intrinsic spin-orbit gap,
$\Delta_{so}\sim 4000 {\rm K}$. Furthermore, in the ferromagnetic phase and
in the vicinity of the Curie temperature, $T_C$, the almost classical $S=5/2$ spins of
the Mn ions fluctuate slowly in time, and therefore, at the time scales and
temperatures  of interest, time reversal is also locally broken, even in the
paramagnetic phase. As a
result,  as we showed already in Ref.~\onlinecite{Gergely}, \gamnas belongs to the {\em unitary}
class.\cite{Arne} Then the $\gamma_t$ plays no role, and
$\gamma_s$ can also be set to $\gamma_s=1$.\cite{Finkelstein,Castellani}
As a result, the scaling of the dimensionless conductance is described by a scaling equation,
\begin{equation}
\frac{d\,\ln g}{dx} = \beta(g)\;,
\label{eq:beta_function}
\end{equation}
where $x=\ln(\xi)$ is a scaling variable with
$\xi=\xi(T)$ a lengthscale, and  due to the simplifications above, the
$\beta$-function depends only on $g$ itself~\cite{Finkelstein,Finkelstein_private}.
In three dimensions, there is a metal-insulator transition
characterized by $\beta\left(g_C \right)= 0$, with $g_C$ the
critical conductance. For $g>g_C$  and $\beta\left(g_C \right) >0$,
the conductor is metallic, and the dimensionless conductance
increases with increasing system size, while $\beta\left(g<g_C
\right) <0$,  and one recovers an insulating state.

Would we know the $\beta$-function, we could compute the resistivity
as follows: Suppose we know the typical dimensionless
conductance $g_0$ at an energy scale $T_0$ and at the corresponding
 microscopic length scale,  $\xi_0= \xi(T_0)$.
Then the resistivity of a large three-dimensional conductor can
be computed by integrating the scaling equation up to a length scale $x=\ln(\xi(T)/\xi_0)$
and  cutting the system to small cubes of size  $\xi=\xi(T)$ to give
\begin{equation}
  \varrho \left (T,H\right) = %C
\frac{h}{e^2}
%\frac{1}{g(\xi)}\frac{\xi}{f\left(\xi/L_T \right)}\;,
\frac{\xi(T)\; g_C}{g(\xi(T),g_0/g_C)}\;.
\label{eq:resistivity}
\end{equation}
%with $C$ a dimensionless constant of the order of unity.
 Note that here we compute the typical conductance.
Corrections due to universal conductance fluctuations can give a
singular contribution,~\cite{Felix}. However, these corrections  are
very small for magnets with a short mean free path such as \gamnas,
and therefore they cannot explain the resistivity anomaly in \gamnas
\cite{Felix}.
%The constant $C$ above is not precisely known, we therefore
%determined it from the experimental data and set it to $C\approx 4$ to
%obtain an optimal fit for  all samples considered.
%We emphasize that, the fact that we were able to fit all samples with a {\em single }
%parameter $C$ is very remarkable, and is an indirect proof of the
%correctness of attributing these anomalies to strong localization
%effects.

To compute $g(\xi)$ we need to know the beta-function.
Unfortunately, while this can and has  been determined numerically
for a non-interacting unitary system \cite{Gergely}, it is
 not known  for interacting electrons.
Here we shall use its  asymptotic form on the metallic side,
 $\beta( g) =1  - {g^{(0)}_C}/{g}$,
with $g^{(0)}_C$ the critical value of the conductance to lowest order in the
epsilon expansion.

Furthermore, we need to know the connection between the scale $\xi(T)$
and the temperature $T$.
This connection can be established by looking at
the pole structure of the diffusion propagator,
$\xi^2(T) \sim D(T) / (T\;z(T))$,  and using the Einstein relation, $\sigma(T) \sim (e^2/h) \; N(0) D(T)$
together with Eq.~\eqref{eq:resistivity} \cite{Castellani,Finkelstein}. However, since
we are interested in the metallic regime, with a good approximation, we can neglect the
energy-dependence of the factor $z(T)$ and set it to $z\equiv 1$. As a result,
we have $N(0)\;  \xi^3 \sim {g(\xi)}/{T}$
\cite{Finkelstein}, which can equally be written as
\be
\left(\frac{\xi}{\xi_0}\right)^3 = \frac{g(\xi/\xi_0, g_0)}{g_0}
\frac{T_0}{T}\;,
\label{eq:xi(T)}
\ee
with $T_0$ the energy scale corresponding to the length scale $\xi_0$.

With the $\beta(g)$ and $\xi(T)$ at hand, we then only need one more ingredient,
the microscopic resistivity, $g_0=g_0(h,t)$, with
$t=T/T_C$ and $h
= g\, \mu_B\, H\, S /T_C$ denoting the dimensionless temperature and
magnetic field, respectively. In the following, we shall assume that
$g_0$ depends exclusively on the  magnetization $m$ of the
sample,
\be g_0 \left(t, h
\right) = g_0(m(t,h))\approx
\tilde g_0
   \left(1+ q\, m^2\left(t,h\right)\right)
\label{eq:g0approx}
\ee
where $\tilde g_0$
is the conductance  of the unpolarized  system. This
approximation is well justified within a mean-field
description of the scattering on spin disorder~\cite{Gergely}, but it also emerges
quite naturally  for  mechanisms where the primary role of the
local magnetic moments is just to polarize the charge carriers and
indirectly influence their  scattering rate~\cite{Brey}. In the present
formalism, however, the precise microscopic origin of the $m$-dependence
of $g_0$ is of secondary importance.

The quadratic term in Eq.~(\ref{eq:g0approx}) is the leading
contribution allowed by time reversal symmetry, and provides a very
good approximation  even in the extreme case of the  infinite
coupling disordered Kondo lattice~\cite{Gergely}. From the fits we found that
the  parameter $q\approx 0.5\div 0.7$ for all samples, in  rough agreement
with the results of \cite{Brey}.

\begin{figure}[tb]
\centering
\includegraphics[width=3.3in,clip]{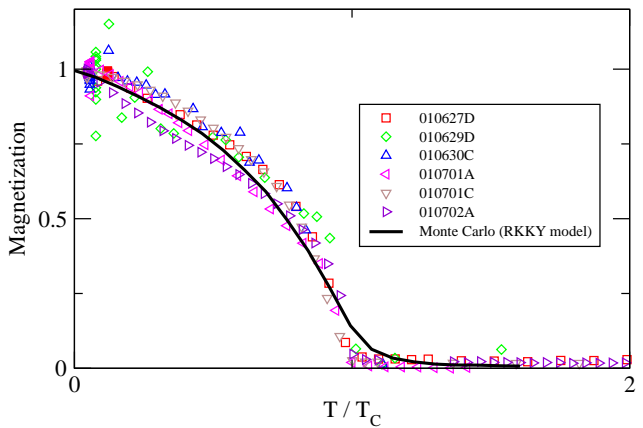}
\caption{ Temperature
dependence of the normalized magnetization in an external
(H=50 Oe) in-plane magnetic field. For each sample the
we rescaled $T$ by the corresponding $T_C$.
The solid line is the result of a Monte Carlo calculation
for a diluted magnetic semiconductor \gamnas by assuming a RKKY interaction
between the localized {\rm Mn} spins.\label{fig:magnetization}}
\end{figure}

In principle, we could now use the experimental $m(t,h)$ curves to compute the
magnetoresistance of a sample.
Unfortunately, experimentally, at large magnetic fields we could not
separate the magnetization of the \gamnas film and that of the
paramagnetic substrate, and only magnetization curves at small magnetic
fields (H = 50 Oe)  were available. Therefore, instead of using the
experimental data,
we determined  the magnetization $m(t,h)$ in a finite field
by performing simulations for a diluted spin
system. To simulate \gamnas we
placed  magnetic ions with a given concentration
at random positions of an  FCC lattice
following the procedure of Ref.\cite{Greg},
We assumed an RKKY interaction between the Mn spins and
computed the magnetization curves $m(t,h)$ by  performing
a Monte Carlo simulation. %While this procedure cannot reproduce the
%precise value of the Curie temperature, $T_C$,
This procedure reproduces the $m(t,h)$ curves, which turn out to be almost concentration-independent
(apart from the limit of very small Mn concentrations), and
fit very nicely the experimentally measured magnetization for $h=0$ too (see Fig.~\ref{fig:magnetization}).
Although these magnetization curves look rather similar to the ones obtained from a simple
mean field approximation, $T_C$ is
suppressed by a factor $\sim 2$  compared to the mean field Curie
temperature due to thermal fluctuations. These thermal fluctuations
being very sensitive to the magnetic field, the
magnetization induced by an external field of $\sim 1\;{\rm T}$ at $T=T_C$ was about a
factor $\sim 1.5$ larger than the mean field estimate. This
increased response was important to obtain the correct position
of the resistivity maximum in finite magnetic fields. Note, however, that the
the curves $m(t,h)$ obtained this way have no free fitting parameters.
\begin{table*}
\caption{\label{tab:table1}Characteristic
parameters of the samples analyzed:
 $x$ is the \rm{Mn} concentration,
$\varrho (T_C)$ is the  resistivity at
 $T_0\equiv T_C$, and $k_F$ stands for the Fermi momentum obtained
by assuming a compensation of $50 \%$. We computed  $k_F l$  from
the Drude formula. We also show  the Fermi wavelength  $\lambda_F$
of the non-magnetic system,  the fitted correlation length at $T_0=T_C$
[$\xi_0$] and at
$1K$ [$\xi(1K)$], respectively,  the
dephasing length   $ \xi^{\rm Drude} (T_C)$ obtained
using the Drude estimate, as described in the text. Finally we list
the fitted values of $\tilde g_0/g_C$.
}
\begin{ruledtabular}
\begin{tabular}{cccccccccc}
Sample number & $x$ & $T_C$ & $\varrho (T_C)$  &$k_F\, l(T_C)$ &
$\lambda_F(T_C)$  &$ \xi^{\rm Drude}(T_C)$
&$ \xi^{\rm fit} (1 K)$ &$ \xi^{\rm fit} (T_C)$ & $\tilde g_0/g_C$  \\
 &(\%) & $ (K)$&$\Omega\, cm$ & & $ (nm)$ &$ (nm)$ &$ (nm)$ &$(nm)$ &\\  \hline
010627D & 1.35 & 42 & $27\times 10^{-3}$  &0.24 & 3.34  & 3.59 &
54.43 & 8.4
& 1.35 \\
010629D & 2.78 & 65 & $10\times 10^{-3}$  &0.51 & 2.64  & 4.38 &
70.94 & 8.8
& 2.85 \\
010630C & 3.91 & 90 & $4.5\times 10^{-3}$ &1.01 & 2.35  & 5.29 &
96.76 & 10.2
& 4.16 \\
010701A & 4.53 & 85 & $11\times 10^{-3}$ &0.39 & 2.24   & 3.43 &
94.03 & 10.2
& 4.01 \\
010701C & 5.87 & 110& $3.6\times 10^{-3}$ &1.10 & 2.04  & 4.95 &
100.68 & 9.6
& 4.76 \\
010702A & 6.68 & 70 & $35\times 10^{-3}$ &0.10 & 1.97   & 1.82 &
73.62 & 8.8
& 1.70 \\
\end{tabular}
\end{ruledtabular}
\end{table*}

We now have all ingredients to compute the magnetoresistance.
The temperature and magnetic field-dependence of the resistivity
then originates from the temperature and magnetic field-dependence of the
microscopic conductance $g_0$ and that of the  scale
$\xi$: The correlation length $\xi$ becomes larger as $T\to 0$, and
therefore the  resistivity
increases. This results ultimately in the low-temperature upturn of
the resistivity and is also responsible for the upturn of the resistivity above
$T_C$. At very low temperatures this results in a $\sim \sqrt{T}$-dependence \cite{Altshuler}.
Entering the ferromagnetic phase, or polarizing the {\rm Mn}
moments with an external field, on the other hand, increases $g_0$,
and hence decreases the resistivity. It is the competition
of these two effects that yields the resistivity
anomaly at $T_C$.

Eqs.~(\ref{eq:resistivity}) and  (\ref{eq:g0approx})  together
with Eq.~(\ref{eq:xi(T)}) provide a full theoretical
description of the magneto-resistance in terms of  three
parameters for every sample, $\xi_0$, $\tilde g_0/g_C$, and the
phenomenological parameter $q$.
Fig.~\ref{fig:resistivity} shows the best fits obtained in this way
for six different samples. For all samples we
 defined  $\xi_0$ and $g_0$ as the scale and dimensionless conductance
at $T_0\equiv  T_C$. The parameters of the
 samples are summarized in Table~\ref{tab:table1}.
Note that the position, the shift and the
amplitude of both the resistivity maxima at $T_C$ as well as that of
the low-temperature anomaly work out very nicely for these samples.
This proves indirectly, that both anomalies are related to the vicinity of the
metal-insulator transition.

In Table~\ref{tab:table1} we also enumerate  the values of $k_Fl$ we
obtain from the resistivity by assuming
a valence hole band of effective mass $m^*=0.45\; m_e$ and
a compensation of $50 \%$. Clearly, the values obtained in this way
are inconsistent with a weakly disordered free electron picture,
but
could naturally be explained through the presence of an impurity
band with an enhanced carrier mass.\cite{Basov}.
% {\bf Pascu: What did you assume for spin
%  degeneracy ?}

Importantly, the fitted values of $\xi$ are smaller than the
thickness of the films, $W\approx 123 \;\rm nm$ even at $T=1K$,
and  they are in good agreement with the values obtained in Refs.~\cite{Weiss}.
These samples are thus three-dimensional from the point of view
of conductance properties down to these temperatures.  Also, $\xi$  remains
larger  than the typical Mn-Mn separation $\sim 1 \rm nm$ over the whole range
of temperatures,  thereby justifying the scaling approach used here.
In Table~\ref{tab:table1} we also included the theoretical estimate
of $\xi^{\rm Drude}(T)= \sqrt{D_{\rm Drude}/T} $ where we estimated
$D_{\rm Drude}$ by using the Drude formula
and the density of
states of a parabolic valence band with a renormalized mass.
The values obtained in this way
do not depend too much on the specific sample, and, apart from an overall
factor,  are in rough
agreement with the values extracted from the fits. This is somewhat
surprising in view of the extremely small values of  $k_F l$. % which seem to be
%inconsistent with a simple disordered valence band picture
%\cite{Basov?,Furdyna}.

In conclusion, we presented here a systematic study of the
resistivity  of various \gamnas samples. We argued that even the
annealed samples are very close to the localization transition and
showed that  the complete magnetic field dependence of the
resistivity anomaly at $T_C$ as well as the low temperature upturn
of the resistivity can be {\em quantitatively } described in terms
of a perturbative scaling theory of localization, combined with
Monte Carlo simulations and a reasonable choice of materials
parameters.

We would like to thank J.K. Furdyna, X. Liu, A. H. MacDonald, J.
Sinova, and T. Jungwirth, but especially A.M. Finkelstein  for
helpful discussions. This research has been supported  by Hungarian
grants OTKA Nos. NF061726, T046267, and T046303, Romanian grant
CNCSIS 2006/1/97 and CNCSIS 2007/1/780. BLS, PS, and NS were
supported by the NSF and the NNIN. B.J. was supported by NSF-NIRT
award No. ECS-0609249, as well as the US. Department of Energy,
Basic Energy Sciences. under Contract No. W-31-109-ENG-38.

%Unused bibitems

%\bibitem{footnote2} At higher temperatures $1/\tau_{\Phi}$ increases  faster than linear.
%\bibitem{altshuler}
%\bibitem{Belitz} D. Belitz and K. I. Wysokinski, Phys. Rev. B{\bf 36}, 9333 (1987).
%\bibitem{Lin} For a review, see J. J. Lin and J. P. Bird, J. Phys. Condens. Matter {\bf 14}, R501 (2002).
%\bibitem{Kravchenko}S. V. Kravchenko, M. P. Sarachik, Rep. Prog. Phys. {\bf 67}, 1 (2004).
%\bibitem{Dietl_localization}   T. Wojtowicz {\em et al.}, Phys. Rev. Lett. {\bf 56}, 2419-2422 (1986).

\begin{references}
\bibitem{Brey}  M. P. L-Sancho and L. Brey, Phys.Rev. B {\bf 68},
113201 (2003); We believe that
that in \gamnas the mechanism discussed here is the primary
candidate to produce the $m$-dependence of $g_0$.
\end{references}
\end{document}